\documentclass[12pt]{article}
\usepackage{epsf}
\title{\bf Strong decays and Adler selfconsistency condition in
two-dimensional QCD}
\author{
Yu.S.Kalashnikova\thanks{e-mail: yulia@vxitep.itep.ru}, A.V.Nefediev
\thanks{e-mail: nefediev@vxitep.itep.ru}}
\date{\it Institute of Theoretical and Experimental Physics,\\
117218, Moscow, Russia}
\newcommand{\be}{\begin{equation}}
\newcommand{\ee}{\end{equation}}
\newcommand{\ra}{\rightarrow}

\newcommand{\vpint}{\int\makebox[0mm][r]{\bf --\hspace*{0.13cm}}}
\newcommand{\ds}{\displaystyle}
\newcommand{\vph}{\varphi}
\begin{document}
\maketitle

\begin{abstract}
Strong mesonic decays are studied in the framework of the 't~Hooft
model for the two-dimensional QCD in the axial gauge. Special attention is payed to the processes with
pions in the final state involved and the low energy theorems have been checked
including the \lq\lq Adler zero".
It is demonstrated explicitly that in the chiral limit any three-meson decay amplitude with at least one
pion vertex vanishes identically for any values of the external momenta. Ward
identities for the dressed vector and axial-vector currents vertices are
identified and their relationship to the pionic vertex are discussed. 
\end{abstract}

It has been recognized that strong decays play the crucial
role in hadron spectroscopy issues. There are two reasons for this. First,
as the distances involved in the light quark sector are large, the
dominant mechanism of strong decay should be nonperturbative, so
that the decay studies deliver knowledge on the picture of confinement.
Another point is that strong hadronic decays are proved to be a rather
important tool in searches for constituent glue. The characteristic decay
selection rules are known which distinguish between $e.g.$ $q\bar q$ and
hybrid mesons \cite{selectionrules}. These selection rules are derived
on the basis of quark models, and the question remains unanswered
of how reliable are the model predictions from the $QCD$ motivated
point of view.

In the constituent quark picture approach the confinement is modeled
by a potential force, and the resulting spectrum describes
the data reasonably well with notable exception, the pion. The chiral
symmetry breaking (CSB) effects lie definitely beyond the scope of a simple
constituent picture, and there is no hope to obtain Goldstone 
boson within any naive quark model. Similarly, decays with pions 
in the final state present problems for pair creation models. For example, 
in the standard $^3P_0$ model \cite{3P0,KI,phasespace} a reasonable fit to 
experimental data is obtained only with the so-called "mock meson"
prescription for the phase space \cite{KI}, while for the correct
relativistic phase space the rate for processes with pions in the final
state is too low \cite{phasespace}. On the contrary, the microscopic model
\cite{Barnes} 
of hadronic decays overestimates the decays with pions for higher 
quarkonia \cite{YuSK&AN}. What is worse,
the naive pair creation models are based on the constituent picture and
do not respect soft pions theorems; in particular, there is no hope to
obtain Adler selfconsistency condition \cite{Adler} for the amplitude 
with pions. Unfortunately, the decays of immediate relevance for hadron
spectroscopy are mostly the ones with pions in the final state 
\cite{AD&YuSK}.

Adler selfconsistency condition follows from the most general
symmetry considerations, and has nothing to do with the particular
mechanism of confinement. Nevertheless, the idea that confinement
and CSB phenomena are interrelated seems to be rather meaningful.
A model was suggested many years ago \cite{Orsay} which connects
confinement and CSB (see also $e.g.$ \cite{port}, where similar ideas
were employed). 
Unfortunately, this model main ingredient, three-dimensional 
oscillator confining force, is not motivated by QCD,  
and among other drawbacks of such an approach is lack of gauge and 
Lorentz invariance. An important step was made in \cite{Sim}, where it was
shown that the same gauge and Lorentz invariant 
nonperturbative gluonic correlators which 
produce area law are responsible for the formation of chiral 
quark condensate. Thus the quark model motivated by such a picture 
should be free of above-mentioned drawbacks and should be able to 
reproduce, {\it inter alia}, all pion properties. Obviously this
quark model is going to be rather unusual, and it is instructive
to study strong decays in some toy exactly solvable theory with
confinement and CSB.  

An example exists of a theory where both confinement and CSB 
are due to the same force, and it is  two-dimensional quantum chromodynamics 
($QCD_2$) in the large $N_C$ limit which was first considered many years 
ago \cite{'tHooft} and still remains popular in studies of various aspects 
of strong interactions. The large $N_C$ limit is important in establishing 
the chiral properties of the theory. Indeed, it is well known that the 
Coleman theorem \cite{Coleman} prohibits CSB for any finite number of 
colours in a two-dimensional theory. Still there is no contradiction with the 
Coleman theorem \cite{Coleman} if one considers the weak coupling regime of 
the theory where 
$m_q\gg g\sim 1/\sqrt{N_C}$, {\it i.e.} the limit of infinite number of colours
is taken first (see {\it e.g.} \cite{Zhitnitskii} for detailed discussion of 
this issue as well as of another phase of the theory which corresponds to the
strong coupling regime $m_q\ll g$ and where the Coleman theorem works at full
scale). 

Most of the studies in $QCD_2$ were performed 
in the light-cone gauge, which considerably simplifies the spectrum 
calculations but yields perturbative vacuum. There exists an
alternative approach based on the Coulomb gauge $A_1 = 0$ 
\cite{Bars&Green}. It appears that the vacuum is nontrivial in the Coulomb 
gauge version, and nonzero quark condensate exists for massless quarks 
\cite{Ming Li}. The latter feature is confirmed by
sum rules approach \cite{Zhitnitskii} in the light-cone gauge.  
 
In the present letter we perform studies of strong hadronic decays
in the $QCD_2$ employing the Coulomb gauge. This choice enables
treating pions on the same footing as other mesons. The latter is in contrast
to light-cone gauge choice, where the pion wave function is very singular,
and one is forced to use sophisticated nonperturbative methods, like
sum rules and operator product expansion, to arrive at reliable results in the
pion physics \cite{Zhitnitskii}.      

Let us review the essentials of $QCD_2$ in the Coulomb gauge (see
\cite{Bars&Green} for the details). 
Our convention for $\gamma$ matrices is $\gamma_0 = \sigma_3$,
$\gamma_1 = i\sigma_2$, $\gamma_5 = \gamma_0\gamma_1$. The large $N_C$
limit implies that $g^2N_C$ remains finite. The gluonic propagator in the 
Coulomb gauge $A_1 = 0$ takes the form
$D_{00}(k_0,k) = - 1/k^2$, and the infrared singularity is regularized
by the principal value prescription yielding the linear confinement. There
are no transverse gluons in two dimensions, and in the large $N_C$ limit
only planar graphs survive, so the only nontrivial one-particle quantity
is the quark Green's function
\be
S(p_0,p)= \frac{1}{\hat p - m - \Sigma(p)}, 
\label{1}
\ee
which can be found from the Schwinger--Dyson equation with the result
\be
\Sigma(p) = [E(p)\cos\theta(p)-m]+\gamma_1[E(p)\sin\theta(p)-p],
\label{2}
\ee
\be
p\cos\theta(p)-m\sin\theta(p)=\frac{\gamma}{2}\vpint\frac{dk}{(p-k)^2}
\sin[\theta(p)-\theta(k)],
\label{3}
\ee
\be
E(p)=m\cos\theta(p)+p\sin\theta(p) + \frac{\gamma}{2}\vpint\frac{dk}{(p-k)^2}
\cos[\theta(p)-\theta(k)],
\label{4}
\ee  
where $\gamma=\frac{g^2N_C}{4\pi}$.
Parameter $\theta(p)$ has the meaning of the Bogoliubov--Valatin angle
describing the rotation from bare to dressed quarks. The obvious
properties of the solutions to the equations (\ref{3}), (\ref{4}) are
$\theta(-p)= -\theta(p)$, $E(-p)=E(p)$, and 
$\theta(p)\mathop{\to}\limits_{p\to\infty}\pi/2$.

The spectrum of the theory is defined from the homogeneous Bethe--Salpeter 
equation for the matrix wave function
\be
\Phi(p,P)=T(p)\left(\frac{1+\gamma_0}{2}\gamma_5\varphi_+(p,P) + 
\frac{1-\gamma_0}{2}\gamma_5\varphi_-(p,P)\right)T^+(P-p)
\label{5}
\ee 
with $T(p)=exp(-\frac{1}{2}\theta(p)\gamma_1)$. The functions
$\varphi_{\pm }(p,P)$ are solutions to the system of equations
\be
\left\{
\begin{array}{c}
[E(p)+E(P-p)-P_0]\vph_+(p,P)\hspace*{8cm}\\
\hspace*{4.2cm}=\gamma\ds\vpint\frac{\ds dk}{\ds (p-k)^2}
\left[C(p,k,P)\vph_+(k,P)-S(p,k,P)\vph_-(k,P)\right]\\
{}\\

[E(p)+E(P-p)+P_0]\vph_-(p,P)\hspace*{8cm}\\
\hspace*{4.2cm}=\gamma\ds\vpint\frac{\ds dk}{\ds (p-k)^2}
\left[C(p,k,P)\vph_-(k,P)-S(p,k,P)\vph_+(k,P)\right],
\end{array}
\right.
\label{6}
\ee
where 
$$
C(p,k,P)=\cos\frac{\theta(p)-\theta(k)}{2}\cos\frac{\theta(P-p)-\theta
(P-k)}{2},
$$
\be
S(p,k,P)=\sin\frac{\theta(p)-\theta(k)}{2}\sin\frac{\theta(P-p)-\theta
(P-k)}{2},
\label{7}
\ee
with $P$ and $p$ being the total meson momentum and the momentum of the quark
respectively.

The system (\ref{6}) describes a meson moving forward in time as a
superposition of the quark-antiquark pair moving forward in time with
the amplitude $\varphi_+$ and backward in time with the amplitude $\varphi_-$.
Such a particle-hole interpretation is supported by the Hamiltonian approach
developed in \cite{Ham}, where the operator $m_n^+(P)$ creating the $n$-th
meson with total momentum $P$ was constructed as a linear combination
of operators creating and annihilating the $q\bar q$ pair. 

A very interesting point concerning Bars--Green system of equations (\ref{6})
is that if it is treated as a matrix integral equation, then its kernel
is not Hermitian \cite{Ham}. The eigenvalues $P_0^n$ are real, but the 
orthonormality and completeness conditions take the form (see \cite{Ham} for
details)
\be
\begin{array}{rcl}
\ds\int\frac{\ds dp}{\ds
2\pi}\left(\vph_+^n(p,P)\vph_+^{m}(p,P)-\vph_-^n(p,P)\vph_-^m(p,P)
\right)&=&\delta_{nm}\\
&&\\
\ds\int\frac{dp}{2\pi}\left(\vph_+^n(p,P)\vph_-^{m}(p,P)-\vph_-^n(p,P)\vph_+^m(p,P)
\right)&=&0
\label{8}
\end{array}
\ee
\be
\begin{array}{rcl}
\ds\sum\limits_{n=0}^{\infty}\left(\vph^n_+(p,P)\vph^n_+(k,P)-\vph^n_-(p,P)
\vph^n_-(k,P)\right)&=&2\pi\delta\left( p-k\right)\nonumber\\
&&\\
\ds\sum\limits_{n=0}^{\infty}\left(\vph^n_+(p,P)\vph^n_-(k,P)-\vph^n_-(p,P)
\vph^n_+(k,P)\right)&=&0\nonumber.
\label{9}
\end{array}
\ee

Solutions of the system (\ref{6}) come in pairs: for each eigenvalue
$P^n_0$ with eigenfunction $(\varphi^n_+,\varphi^n_-)$ there exists another
eigenvalue $-P^n_0$ with eigenfunction $(\varphi^n_-,\varphi^n_+)$. With
this symmetry only positive eigenvalues enter the forms (\ref{8}), (\ref{9}).

The system (\ref{6}) is solved numerically in \cite{Birse}, and it is shown
that the $\varphi_-$ component is small $i)$ if the quark mass is large 
and $ii)$ for higher excited states. In both cases quark potential
model with local linear confinement serves as a good approximation. 
Besides that the $\varphi_-$ component 
dies out with the increase of the total mesonic momentum $P$: in the
infinite momentum frame ($P \ra \infty$) it is zero and the equation
for the $\varphi_+$ reproduces the 't~Hooft equation \cite{'tHooft}
after an appropriate rescaling, as it was shown in \cite{Bars&Green}. 

The chiral condensate is calculated straightforwardly from equation (\ref{1}):
\be
\langle vac|q\bar q|vac\rangle=-\frac{N_C}{2\pi}\int_{-\infty}^{+\infty}
dk\cos\theta(k).
\ee

It is shown in \cite{Ming Li} that the gap equation (\ref{2}) has 
nontrivial solution in the chiral limit $m=0$, and the chiral condensate
(10) does not vanish with this solution. It means that the Goldstone mode
exists in the spectrum, and its explicit form was found in \cite{Ham}:
\be
\varphi^{\pi}_{\pm}(p,P) = \sqrt{\frac{\pi}{2P}}\left(\cos\frac{\theta(P-p)-\theta(p)}{2}
\pm \sin\frac{\theta(P-p)+\theta(p)}{2}\right) 
\label{11}
\ee
for $P>0$ and $P_0=\sqrt{P^2}$ ($\varphi_-^{\pi}\leftrightarrow\varphi_+^{\pi}$
for $P<0$).  For $P=0$ $\varphi_+^{\pi}(p,0) = \varphi_-^{\pi}(p,0)$, pion
spends half of time  in backward motion of the pair, and such a function 
has zero norm, as it should be for the massless particle at rest.
In the opposite limiting case $P\ra\infty$ the backward motion part
dies out, and
\be
\varphi_+^{\pi}(p,P)= \sqrt{\frac{2\pi}{P}},\quad 0\leq p\leq P,
\label{12}
\ee
coinciding with the Goldstone mode of the 't~Hooft equation. Form (\ref{12})
is the main reason to consider the Coulomb gauge version of the theory, as
all nontrivial content of the wave function (\ref{12}) is concentrated in 
the boundary
regions $x \ra 0$ and $x \ra 1$ ($x = p/P$). The same is true of
course for the $QCD_2$ quantized at the light-cone \cite{Callan}. 
Quantities like chiral condensate do not come out trivially with
such a singular wave function \cite{Zhitnitskii}.
\begin{figure}[t]
\hspace*{1cm}
\epsfxsize=14cm
\epsfbox{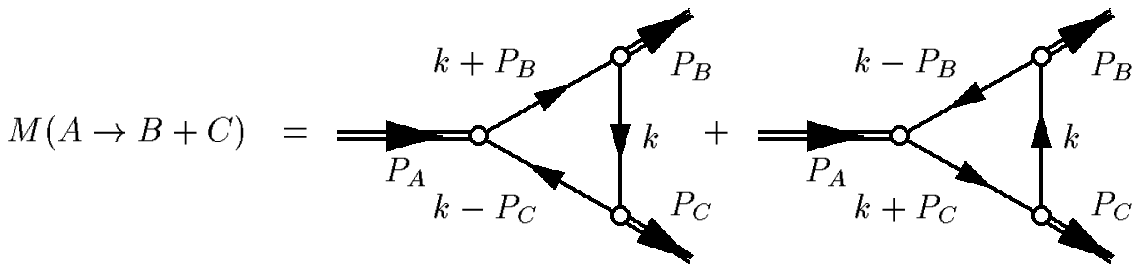}
\caption{Graphical representation of the decay amplitude (\ref{13}).}
\end{figure}

Now we are in the position to calculate the strong decay amplitude
(Fig.1), which is $O(1/\sqrt{N_C})$.
The answer reads 
$$
M(A \ra B+C)=\hspace*{7cm}
$$
\be
-\frac{i\gamma^3}{\sqrt{N_C}}\int \frac{d^2k}{(2\pi)^2} Sp(\Gamma_A(k+P_B,P_A)S(k-P_C)\bar \Gamma_C(k,P_C)
S(k)\bar \Gamma_B(k+P_B,P_B)S(k+P_B))
\label{13}
\ee
$$
\hspace*{7cm}+(B \leftrightarrow C),
$$
where the meson-quark-antiquark vertex $\Gamma_M$ can be expressed in terms of the
Bethe--Salpeter wave function (\ref{5})
\be
\Gamma_M(p,P)=\int\frac{dk}{2\pi}\gamma_0\frac{\Phi_M(k,P)}{(p-k)^2}\gamma_0,
\label{14}
\ee
and $\bar \Gamma_M(p,P)=\gamma_0\Gamma_M^+(p,P)\gamma_0$. As the vertex
(\ref{14})
does not depend on $p_0$, the integration along the loop in (\ref{13}) is trivial,
so one arrives at the following final expression for the decay amplitude in 
the initial meson rest frame $P_A=0$, $P_B=-P_C=p$:
\be
M(A \ra B+C)=
\label{15}
\ee
$$
\frac{\gamma}{\sqrt{N_C}}\int\frac{dkdq}{(q-k)^2}\left\{\right.
-\varphi_-^A(k+p,0)\varphi_-^B(k+p,0)[c(-p,q,k)\varphi_+^C(q,-p)+s(-p,q,k)\varphi_{-C}]
$$
$$
-\varphi_+^A(k+p,0)\varphi_+^C(k,-p)[c(p,q+p,k+p)\varphi_+^B(q+p,p)+s(p,q+p,k+p)\varphi_-^B(q+p,p)]
$$
$$
-\varphi_+^C(k,-p)\varphi_-^B(k+p,p)[s(0,q+p,k+p)\varphi_+^A(q+p,0)+c(0,q+p,k+p)\varphi_-^A(q+p,0)]
$$
$$
+\varphi_-^C(k,-p)\varphi_+^B(k+p,p)[c(0,q+p,k+p)\varphi_+^A(q+p,0)+s(0,q+p,k+p)\varphi_-^A(q+p,0)]
$$
$$
+\varphi_-^A(k+p,0)\varphi_-^C(k,-p)[s(p,q+p,k+p)\varphi_+^B(q+p,p)+c(p,q+p,k+p)\varphi_-^B(q+p,p)]
$$
$$
+\varphi_+^A(k+p,0)\varphi_+^B(k+p,p)[s(-p,q,k)\varphi_+^C(q,-p)+c(-p,q,k)\varphi_-^C(q,-p)]\left.\right\}
$$
$$
+(B \leftrightarrow C, p \leftrightarrow -p),
$$
where 
$$
c(p,q,k)=\cos\frac{\theta(k)-\theta(q)}{2}\sin\frac{\theta(p-k)-\theta(p-q)}{2},
$$
$$
s(p,q,k)=\sin\frac{\theta(k)-\theta(q)}{2}\cos\frac{\theta(p-k)-\theta(p-q)}{2}.
$$

Amplitude (\ref{15}) could be easily obtained from the Hamiltonian approach
developed in \cite{Ham}, though further calculations appear more transparent
using the matrix wave function (\ref{5}) and the matrix form of the bound state
equation (\ref{6}). Nevertheless, equivalence of both approaches can be established at any
intermediate step.

If one neglects the backward motion contributions in the amplitude (\ref{15})
and inserts the nonrelativistic values of the angle $\theta$
($\cos\theta(k)=1, \sin\theta(k)=k/m$) then it reproduces the
standard quark model decay amplitude due to OGE Coulomb interaction
\cite{Barnes} adapted to the two-dimensional case. It is clear, however, that substitution
of the nonrelativistic angle is not justified for kinematical reasons.

The six-term form (\ref{15}) was anticipated in \cite{Swanson}, where the
role of backward motion in decay was discussed for the first time. It was argued
there that a simple procedure could be used as a crude approximation.
Namely, one neglects the backward motion pieces for all mesons but pions.
Then a simple graph counting would give the net result: amplitudes with a
pion in the final state are multiplied by an overall factor 2 over the naive 
quark model result, and those with two pions are multiplied by 3. 
Obviously this prescription of constructive interference does not solve the 
Adler selfconsistency problem.

It appears that in the two-dimensional amplitude (\ref{15}) the interference between the six
terms is destructive: {\bf the total amplitude vanishes in the chiral
limit, if at least one of the final state mesons is the pion}. Of course, one 
can perform explicit calculations inserting the pion w.f. (\ref{11}) and
making use of the equations (\ref{3}), (\ref{4}) and (\ref{6}).
There exists, however, a simple and elegant expression for the
pion-quark-antiquark vertex in the limit $m=0$:
\be
\Gamma_{\pi}(p,P)=S^{-1}(p)(1+\gamma_5)-(1-\gamma_5)S^{-1}(p-P).
\label{16}
\ee

Inserting it into equation (13) trivially yields $M(A \ra \pi+C)=0$ in the
chiral limit.

Expression (\ref{16}) for the pion vertex follows from rather general
considerations and is connected with the vector and axial-vector currents
conservation.  \begin{figure}[t]
\hspace*{3cm}
\epsfxsize=10cm
\epsfbox{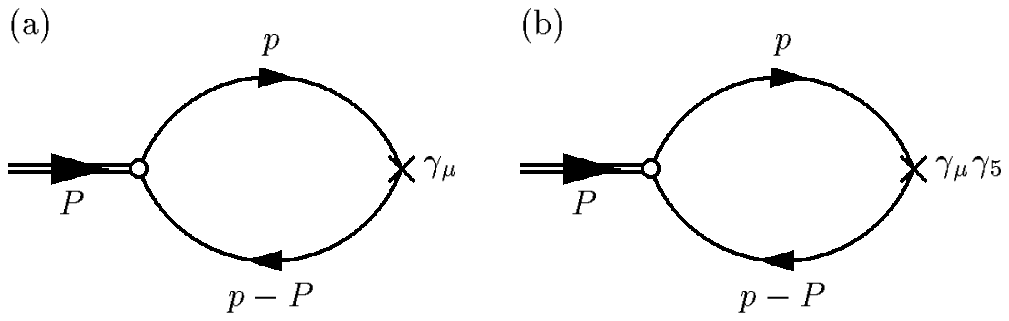}
\caption{The meson-vector (a) and meson-axial-vector (b) current couplings.}
\end{figure}

The meson-vector current coupling  
$V^M_{\mu}(P)=\langle vac|\bar q\gamma_{\mu} q|M,P\rangle$ can be calculated as
(Fig.2a) 
\be
V^M_{\mu}=i\gamma\sqrt{N_C}\int\frac{d^2p}{(2\pi)^2}Sp(S(p-P)\gamma_{\mu}S(p)\Gamma_M(p,P))= 
\sqrt{N_C}\int \frac{dp}{2\pi} Sp(\gamma_{\mu}\Phi_M(p,P)).
\label{17}
\ee

Substituting expression (\ref{5}), one finds
$$
V^M_0
=-\sqrt{N_C}\int\frac{dp}{2\pi}[\varphi^M_+(p,P)+\varphi^M_-(p,P)]\sin\frac{\theta(P-p)
+\theta(p)}{2},
$$
\be
V^M =
-\sqrt{N_C}\int\frac{dp}{2\pi}[\varphi^M_+(p,P)-\varphi^M_-(p,P)]\cos\frac{\theta(P-p)
-\theta(p)}{2}.
\label{18}
\ee

To demonstrate the vector current conservation we write down the equation
for the matrix wave function (\ref{5}) as 
$$
P_0^M\Phi_M(p,P)=(\gamma_5p+\gamma_0m)\Phi_M(p,P)-\Phi_M(p,P)(\gamma_5(P-p)+\gamma_0m)
\hspace*{5cm}
$$
\be
+\gamma\int\frac{dk}{(p-k)^2}\left\{\Lambda_+(k)\Phi_M(p,P)\Lambda_-(P-k)-
\Lambda_+(p)\Phi_M(k,P)\Lambda_-(P-p)\right.
\label{19}
\ee
$$
\hspace*{5cm}\left.-\Lambda_-(k)\Phi_M(p,P)\Lambda_+(P-k)+\Lambda_-(p)\Phi_M(k,P)\Lambda_+(P-p)\right\},
$$
where projectors are defined as $\Lambda_{\pm}(p)=T(p)\frac{1\pm\gamma_0}{2}T^+(p)$.

Then, integrating this equation over $p$, multiplying by $\gamma_0$ and
taking trace one gets
\be
P_0^MV^M_0-PV^M=0,
\ee
$i.e.$ the vector current is conserved. 

Similarly, for the axial-vector current matrix element
$A^M_{\mu}(P)=\langle vac|\bar q\gamma_{\mu}\gamma_5 q|M,P\rangle$ one can find 
(Fig.2b)
\be
P_0^MA^M_0-PA^M =-2m\sqrt{N_C}\int\frac{dk}{2\pi}Sp(\Phi),
\ee
that yields axial-vector current conservation for $m=0$. Slightly relaxing the 
chiral limit one gets the pionic decay constant $f_{\pi}=\sqrt\frac{N_C}{\pi}$ 
and the Gell-Mann--Oaks--Renner relation \cite{GMOR}
\be
f^2_{\pi}M^2_{\pi}=-2m\langle\bar{q}q\rangle
\ee 
(see also \cite{Zhitnitskii,Ham}). It is instructive to mention  
that in the chiral limit vector current has nonzero coupling 
only to the pion: the vector current matrix elements 
(\ref{18}) contain the meson wave functions integrated with the pionic one
(\ref{11}), 
so the corresponding decay constant vanishes due to orthogonality conditions
(\ref{8}).
(We remind here that the pion does couple to the vector current in
1+1!). As the axial-vector current in 1+1
is dual to the vector current, the axial decay constants $f_M$ 
for excited mesons vanish in the chiral limit too.  
\begin{figure}[t]
\hspace*{5.2cm}
\epsfxsize=6cm
\epsfbox{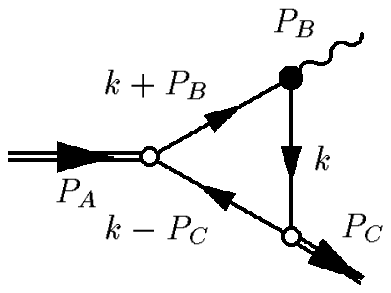}
\caption{Graphical representation of the mesonic form factor.}
\end{figure}

Calculations of mesonic form factors (Fig.3) are more involved, and require 
knowledge of the full quark-antiquark scattering amplitude 
$\Gamma_{ik,lm}(p_{\mu},k_{\mu},P_{\mu})$,
$$
\Gamma_{ik,lm}(p_{\mu},k_{\mu},P_{\mu}) = \frac{2\pi i\gamma}{(p-k)^2}(\gamma_0)_{il}
(\gamma_0)_{km}
-i\gamma^2\sum_M\frac{1}{P_0-P^M_0}(\Gamma_M(p,P))_{im}(\bar
\Gamma_M(k,P)_{kl}, 
$$
\be
+i\gamma^2\sum_M\frac{1}{P_0+P^M_0}(\gamma_0\bar \Gamma_M(P-p)\gamma_0)_{im}  
(\gamma_0\Gamma_M(P-k,P)\gamma_0)_{kl}.
\ee
which is the solution of the inhomogeneous Bethe--Salpeter equation.

The dressed vector current-quark-antiquark vertex $v_{\mu}$ (Fig.4) 
is then given by
$$
v_{\mu}(p,P_{\mu})=i\gamma_{\mu}+i\gamma\sum_M \frac{1}{P_0-P^M_0}\int
\frac{dq}{2\pi} Sp[\gamma_{\mu}\Phi_M(q,P)]\bar \Gamma_M(p,P)
$$
\be
-i\gamma\sum_M \frac{1}{P_0+P^M_0}\int \frac{dq}{2\pi}
Sp[\gamma_{\mu}\Phi_M^+(P-q,P)] \gamma_0\Gamma_M(P-p,P)\gamma_0,
\label{24}
\ee
where $P$ is the total momentum of the quark-antiquark pair, and $p$ is
the momentum of the quark. One easily recognizes the vector current
couplings (\ref{17}) entering (\ref{24}), so that 
in the chiral limit only pion contributes
into $v_{\mu}$, displaying a sort of "vector dominance". The same takes
place with the axial-vector current-quark-antiquark vertex.

Using the equations (\ref{1}), (\ref{2}) and (\ref{5}) together with the completeness condition
(\ref{9}) allows to recast the vector current divergence into the Ward
identity form (similar expression in the light--cone gauge was derived
in \cite{Einhorn}).
\be
-iP_{\mu}v_{\mu}(p,P)=S^{-1}(p)-S^{-1}(p-P),
\label{25}
\ee
which makes vector current conservation manifest for the form factors:
\be
Q_{\mu}\langle M,P|v_{\mu}|M',P'\rangle=0,\quad Q_{\mu}=P_{\mu}-P'_{\mu}.
\ee

Similarly, the axial-vector current divergence in the chiral limit is
\be
-iP_{\mu}a_{\mu}(p,P) = S^{-1}(p)\gamma_5+\gamma_5S^{-1}(p-P),
\label{27}
\ee
and
\be
Q_{\mu}\langle M,P|a_{\mu}|M',P'\rangle=0,\quad Q_{\mu}=P_{\mu}-P'_{\mu}.
\ee

The pion-quark-antiquark vertex (\ref{16}) is nothing but a linear 
combination of the vector and the axial-vector currents divergences (\ref{25})
and (\ref{27}) 
\be
\Gamma_{\pi}(p,P)=-iP_{\mu}v_{\mu}(p,P)-iP_{\mu}a_{\mu}(p,P).
\label{29}
\ee

Relation (\ref{29}) plays the role of PCAC hypothesis in 1+1: as pion
couples both to vector and axial-vector currents, the form (\ref{29})
is inevitable in 1+1 then.
\begin{figure}[t]
\hspace*{1cm}
\epsfxsize=14cm
\epsfbox{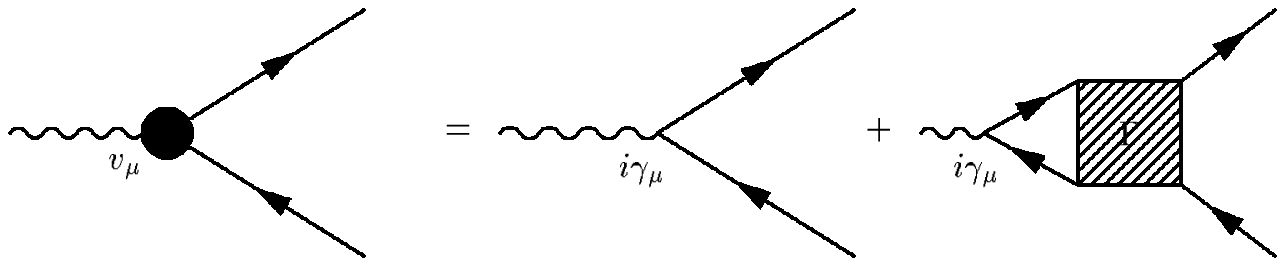}
\caption{Graphical representation of the dressed vector current-quark-antiquark
vertex.}  
\end{figure}

In conclusion, let us briefly recall the results reported in the present 
letter. In the previous publication \cite{Ham} we developed a Hamiltonian 
approach to the two-dimensional QCD in the axial gauge and demonstrated how 
the Hamiltonian of the model could be diagonalized in the mesonic sector.
One of the key features of the above Hamiltonian is the distorted norm 
(\ref{8}) plugged into it which allows description of the massless $q\bar q$
states, chiral pions, on equal footing with the rest of higher massive 
excitations. Chiral properties of the model were reproduced and discussed.
In the meantime a crucial role of the backward in time motion of the 
$q\bar q$ pair inside meson, extremely important for the pion, was anticipated 
but not discussed in detail. In the present letter we return to this 
issue and investigate the role played by pions in the strong hadronic
decays and in providing the low energy theorems like \lq\lq Adler zero" 
selfconsistence condition. We use an effective diagrammatic techniques
involving dressed meson-quark-antiquark vertex, quark and mesonic propagators 
and the full four-quark scattering amplitude. Any hadronic process can be 
formulated diagrammatically using the above ingredients. The three-meson 
amplitude is studied in detail in the suggested approach and a strict result
is obtained stating that the interference between different parts of this 
amplitude which are due to $\varphi_+$ and $\varphi_-$ components of the pion
is completely destructive, so that the total amplitude vanishes for any 
pion momentum, not only in the limit $p_{\pi}\to 0$. The latter result seems 
quite natural if one takes into account that the pionic decay constant $f_{\pi}$ is
dimensionless in 1+1 and does not define any 
scale for soft pions then. 

Developing similar approach for the four-dimensional QCD, which could shed 
light on many puzzles of hadronic spectroscopy, is still an open challenge.
In the model \cite{Orsay} the interaction  is described by the time component
of a vector force, yielding the gap equation similar to (\ref{3}), (\ref{4})
and the Bethe--Salpeter equation similar to (\ref{19}). The existence of a
chiral-noninvariant solution of the gap equation implies the existence of
a Goldstone boson. Axial-vector current is conserved in the chiral limit of
the model, and all the relations of current algebra are satisfied. However,
the model is not covariant, that prevents from proceeding further along the
lines presented here. We point out once more that the approach suggested
in \cite{Sim} is rather promising: the gauge invariant confining interaction
employed in \cite{Sim} is Lorentz covariant by construction, respects the area
law and is responsible for the chiral condensate formation. We expect that the
quark model derived within such a formalism could be able to reproduce all the
nice features described above and, being fully covariant, could allow to 
calculate hadronic observables. 

\bigskip

The authors would like to thank B.L.Ioffe for enlighting discussions.
Financial support of RFFI grants 00-02-17836 and 00-15-96786 and
INTAS-RFFI grant IR-97-232 is gratefully acknowledged.
  
\end{document}